# Visualization of reaction chemistry in W-KClO$_4$-BaCrO$_4$ delay mixtures via a Sestak-Berggren model based isoconversional method

Youngjoon Han[1†], Soyeon Kim[1†], Soyun Joo[1], Chungik Oh[1], Hojun Lee[1], Chi Hao Liow[1], Moon Soo Park[2], Dong Hyeon Baek[2], Seungbum Hong[1,3*]

[1]Department of Materials Science and Engineering, Korea Advanced Institute of Science and Technology, Daejeon, 34141, Republic of Korea

[2]Poongsan Defense R&D Institute, Gyeongju-si, 38026, Republic of Korea

[3]KAIST Institute for NanoCentury, KAIST, Daejeon, 34141, Republic of Korea

[†]These authors contributed equally in completing the manuscript

[*]Author to whom correspondence should be addressed: seungbum@kaist.ac.kr

## ABSTRACT

The combustion delay mixture of tungsten (W), potassium perchlorate (KClO$_4$), and barium chromate (BaCrO$_4$), also known as the WKB mixture, has long been considered to be an integral part of military-grade ammunition. Despite its long history, however, their progressive reaction dynamics remains a question mark, especially due to the complex nature of their combustion reaction. As opposed to a one-step oxidation commonly observed in conventional combustions, the WKB mixture is associated with a multibody reaction between its solid-state components. To this end, the emergence of three combustion peaks, which we corresponded with disparate chemical reactions, was observed using thermogravimetric analysis on two separate WKB mixtures with differing mixture ratios. We applied the stepwise isoconversional method on each of the peaks to match the combustion chemistry it represents to the Sestak-Berggren model and computed the conceptual activation energy. Further plotting the logarithmic pre-exponential factor as a function of the reaction progress, we demonstrate a method of using the plot as an intuitive tool to understand the dynamics of individual reactions that compose multi-step chemical reactions. Our study provides a systematic approach in visualizing the reaction chemistry, thereby strengthening the analytical arsenal against reaction dynamics of combustion compounds in general.

# 1. Introduction

## 1.1 Combustion characterization of the WKB mixture

The WKB delay mixture, which consists of tungsten(W), potassium perchlorate ($KClO_4$), and barium chromate ($BaCrO_4$), is a chemical product utilized to control the specific time delay between two ignition events with high repeatability and reliability [1,2]. While the control of combustion characteristics of these mixtures is integral to realize precise and repeatable delay times, the multi-step reactions and burning mechanisms involved in pyrotechnic mixtures are still not fully understood to this day. This is mainly due to their complex physical and chemical processes, which in case of delay mixtures involve diverse phase reactions such as solid-solid reactions, solid-liquid reactions involving one molten component, and some solid-gas reactions [3]. According to the literature, it is known that the reaction progress of the WKB mixture occurs according to the following chemical reactions:

$$W + 3/8 KClO_4 + BaCrO_4 \rightarrow WO_3 + 3/8 KCl + 4/8 Cr_2O_3 + BaO + 508\ kJ \qquad (1)$$

$$BaO + WO_3 \rightarrow BaWO_4 + 307\ kJ \qquad (2)$$

Despite such proposed difficulties, past studies have improved the understanding of the reaction by controlling various factors that can affect the combustion reaction of WKB mixtures and analyzing how each factor affects this reaction [3–6]. Shachar and Gany studied the burning rate and effective activation energy of different delay mixtures experimentally by using various mixture compositions and tungsten particle sizes [7]. Nakamura et al. reported that the burning rate of $W/KClO_4/BaCrO_4$ delay composition linearly increased with pressure, and significantly increased with higher metal content [6]. However, for the purposes of controlling the delay time and various reaction conditions, the results of research conducted while experimentally controlling one factor cannot but be utilized indirectly.

In this paper, to gain deeper insight into the reaction mechanism of complex reactions, we devised an isoconversional method of measuring the activation energy as a function of the reaction progress for reactions proceeding at disparate temperature ranges using the kinetic parameters extracted from a conventional simultaneous thermogravimetric analysis and differential scanning calorimetry (TG-DSC) measurement. Isoconversional methods can afford an evaluation of the effective activation energy $E_a$, without assuming any particular form of the reaction model, where the change in the $E_a$ variation can generally be associated with a change in the reaction mechanism or the rate-limiting step of the overall reaction [8]. Based on the isoconversional method, Kanagaraj et al showed that accelerated hygrothermal aging affects the performance parameters of tungsten based pyrotechnic delay compositions [1]. However, the utilized model only considers the dependence of the overall reaction on the concentration of the final product, whereas reaction in mixtures rely more on the concentration of the transitory reactants rather than the final products from the end of reactions. For further research, we processed the obtained data from the isoconversional method to visualize the reaction into a color gradient plot, based on the Sestak-Berggren model. We hereby demonstrate a method of reaction chemistry visualization, as a way to intuitively understand the dynamics of individual reactions that compose multi-step chemical reactions.

## 1.2 Theoretical Part

The kinetics of solid-state reactions is commonly described by the product of two well-established functions: the rate constant k, which is dependent on temperature T, and the conversion function f, which is dependent on reaction progress α [9]. By applying the Arrhenius equation, we may describe such relation as follows:

$$d\alpha/dt = k(T) f(\alpha) = A_0 \exp(-E_a/RT) f(\alpha) \qquad (3)$$

where dα/dt is the differential reaction rate, $A_0$ is the pre-exponential factor, R is the universal gas constant, and $E_a$ is the activation energy. While the expression for f(α) may vary depending on the selected physical model, most combustion reactions, for instance, that of gas fuel, can be modeled after the collision theory [10,11], where its reaction dynamics can be explained as a one-step process. In such case, the reaction rate is directly proportional to the probability of collision between the reactants and solely depends on the amount of remaining reactant *i.e.* the reaction progress [12–14]. The more time has elapsed, the less reactant would be remaining, and therefore the lower reaction rate. This is represented by the expression below, where the conversion function f is represented by a polynomial expression of α, and the reaction order of the chemical process is given by the maximal exponent n. Here, the $A_0 f(\alpha)$ term of Equation (3) has been aptly replaced by the collision factor $A(1 - \alpha)^n$.

$$d\alpha/dt = A \exp(-E_a/RT) (1 - \alpha)^n \qquad (4)$$

The parameter n can be loosely interpreted as the number of molecules participating in the collisional reaction and holds significant implications in kinetics.

Unlike such reactions, however, the combustion reaction of powdered reactants presents further complications. While the combustion itself can be modeled by the interaction between the contenders, the reaction is further limited by the nucleation and growth of the product, which also retains a solidus phase. Thus, we must consider both the probability of contact between the solid reactants, as well as the grain growth of the solid product. Here, we adopted the Sestak-Berggren (SB) model [15–17] to represent the reaction, with the reaction rate given as follows:

$$d\alpha/dt = A \exp(-E_a/RT) \alpha^m (1 - \alpha)^n \qquad (5)$$

The added term $\alpha^m$ represents how the more the reaction proceeds, the more the nucleus of the product phase permeates the solid matrix, which acts as the epicenter of further solid-state reaction. Thus, the auto-catalytic nature of the combustion reaction of the WKB mixture is well illustrated by the SB model. Furthermore, applying the logarithm on either side of Equation (5), we see that the logarithm of the reaction rate is linearly proportional to the inverse of the temperature, providing further ground for a quantitative analysis, as given below:

$$\ln(d\alpha/dt) = \ln(A) + m \cdot \ln(\alpha) + n \cdot \ln(1 - \alpha) - E_a/RT \qquad (6)$$

The significance and relation between the three terms that comprise the intercept, ln(A), m·ln(α) and n·ln(1–α) are further discussed throughout the progress of this paper. Hereon, we will refer to these terms as the logarithmic pre-exponential factor, the nucleation factor, and the collision factor, respectively.

## 2. Experimental Section

### 2.1 Combustion characterization of the WKB mixture

The combustion of the WKB mixture and its components have been investigated by simultaneous thermogravimetry and differential scanning calorimetry (TG-DSC, LABSYS Evo, Setaram). The ratio of components in the WKB mixture was adjusted to obtain the target delay time. The TG-DSC graphs were obtained in an Ar chamber for the temperature range of RT to 700°C at a heating rate of 10°C/min and an initial sample weight of 10-15 mg using an alumina crucible. TG-DSC results of the components (W, $KClO_4$, $BaCrO_4$) were used to identify the concurrent peaks that appeared from that of the WKB mixture. We additionally conducted a multiple heating rates DSC at five different heating rates of 2, 5, 10, 15, and 20°C/min, with the same temperature range and experimental conditions as previously indicated.

### 2.2 Isoconversion of the DSC data

The DSC results were further analyzed using the isoconversional method [18–21] for a reliable analysis of the reaction kinetics of the delay mixture. We tracked the reaction progress α and the reaction rate dα/dt of each peak during their respective combustion process by utilizing the ratios between the heat released up to or at a certain time frame and the total heat released throughout the combustion in question, as described in the following formula:

$$\alpha(t) = \frac{\int_{T_i}^{T} q \cdot dT/\beta}{\int_{T_i}^{T_f} q \cdot dT/\beta} \qquad d\alpha(t)/dt = \frac{q(t)}{\int_{T_i}^{T_f} q \cdot dT/\beta} \qquad (7)$$

where q = q(t) is the heat flow measured at time t, T is the temperature of the DSC chamber at time t, $T_i$/$T_f$ is the temperature at which the combustion begins/terminates, and β is the heating rate (°C/s) used in the DSC measurement. $T_i$ was taken from the onset of deviation from the baseline, while $T_f$ was established by adjusting the temperature range such that the integrated area could most closely approximate the reported heat flow of the corresponding peak. Based on the DSC results conducted with the heating rate of 10°C/min of a designated WKB mixture, we computed the reaction rate for fixed values of α from 0.01 to 0.99 with 0.01 increments, for each combustion peak observed (Peak 1, 2, 3, shown in Fig. 1(d)). This was repeated for the DSC data measured at all heating rates to establish individual 1/T vs. logarithmic reaction rate plots for the fixed values of α (Equation 6). We conducted the same analysis for two different mixture samples (Type A and B), so as to observe a consistent trend and further individualize the pyrotechnic behavior of the distinct DSC peaks identified in the initial simultaneous TG-DSC analysis.

### 2.3 Visualization of the Reaction Dynamics

We interpreted the DSC data measured at five different heating rates with regards to the SB model (Equation 5) with the following three steps. First, the logarithm of the reaction rate and the inverse of the corresponding system temperature for fixed values of α, iterated for all extant heating rates, was linearly interpolated to calculate the activation energy, as the slope represents the activation energy divided by the gas constant. Next, we used the said method to calculate the sum of the logarithmic pre-exponential factor, the nucleation factor, and the collision factor at each α, which is acquired by extending the interpolation and reading the y-intercept. These two steps were repeated for disparate values of α to construct a map of the activation energy as a function of the reaction progress α and calculate the mean of the y-intercept values. Finally, we extrapolated the mean y-intercept as a function of ln α and ln (1 - α), and using the method of least-squares, fit the SB model to compute the values of reaction parameters m and n.

## 3. Results and Discussion

### 3.1 Combustion characterization of the WKB mixture

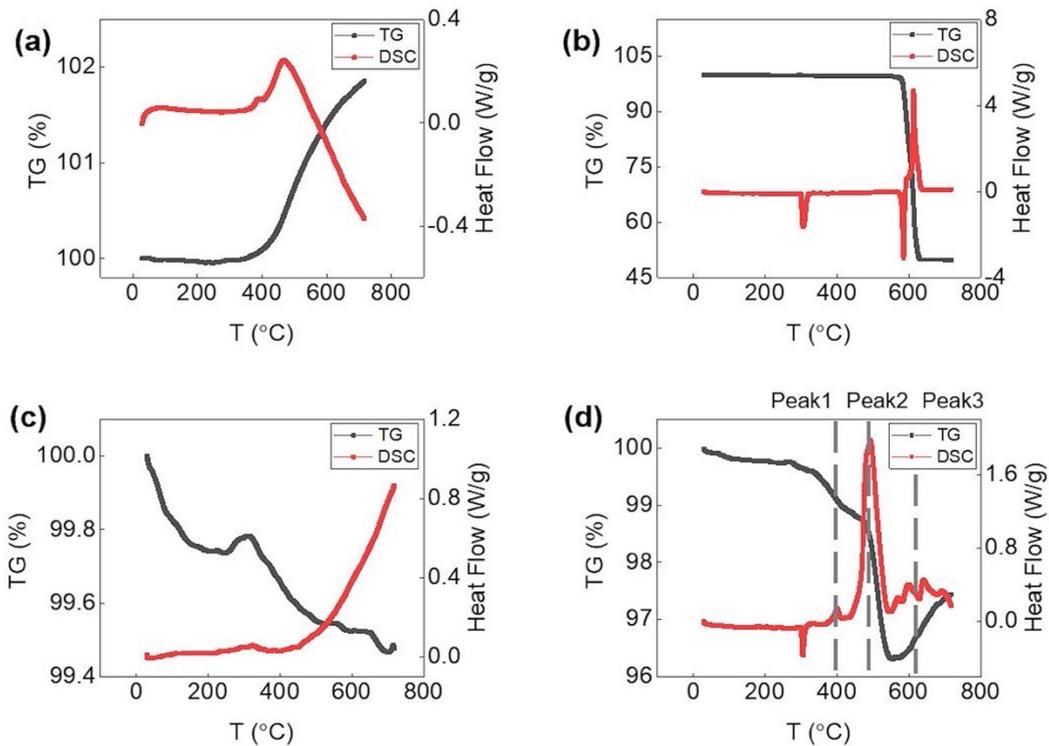

**Fig. 1.** TG-DSC curves for (a) W(b) KClO$_4$, (c) BaCrO$_4$, and (d) WKB mixture, obtained for heating rate of 10°C/min.

Fig. 1(a), (b), and (c) shows the TG-DSC curves for the individual components W, KClO$_4$ and BaCrO$_4$, respectively. Fig. 1(a) shows that W goes through thermal oxidation, with the exothermic reaction hitting its peak in the temperature range between 460°C and 470°C. Although no apparent saturation of mass is observed, there is a total mass increment of ~1.5%. KClO$_4$ showed two endothermic peaks representing its solidus-to-solidus (at 305°C) and solidus-to-liquidus (at 586°C) phase transition, immediately followed by an exothermic dissociation reaction [22], as shown in Fig. 1(b). As the primary oxidant that releases two oxygen molecules within the mixture, larger compositions of KClO$_4$ is associated with greater amounts of heat energy emitted [6]. A total mass change of ~50% shows that all oxygen atoms were detached from the salt (KClO$_4$ → KCl (74.5513 u) + 2O$_2$ (63.996 u) ↑ ), where the reaction is accompanied by partial fusion (KClO$_4$ (s) → KClO$_4$ (l)). While exhibiting a slower combustion rate when compared to KClO$_4$, BaCrO$_4$ also acts as an oxidant within the mixture, where its decomposition reaction subsequently follows that of KClO$_4$ [23]. As BaCrO$_4$ experiences thermal decomposition at 1400°C, only a part of the peak can be observed within the temperature range restricted by limitations of the equipment (Fig. 1(c)). For a higher measuring temperature range of up to 1,500°C, the entire exothermic peak is expected to appear.

Based on the TG-DSC results of each component, we were able to distinguish the combustion peaks of the WKB mixture, represented as Peaks 1, 2, and 3 as shown in Fig. 1(d). While it is easy to deduce by comparison that the endothermic peak at 305°C results from the phase transition of KClO$_4$, the expected reactions of the remaining three exothermic peaks of the WKB mixture stand as follows. Peak 1 (400°C) is possibly a satellite peak of Peak 3, most probably induced by W oxidation. Peak 2 (490°C) is induced by the reaction between W and KClO$_4$, where the thermal decomposition of KClO$_4$ provides the oxygen consumed in the process. Peak 3 (640°C) is a multi-step process that involves the residual W reacting with the remaining BaCrO$_4$.

## 3.2 Isoconversion of the DSC data

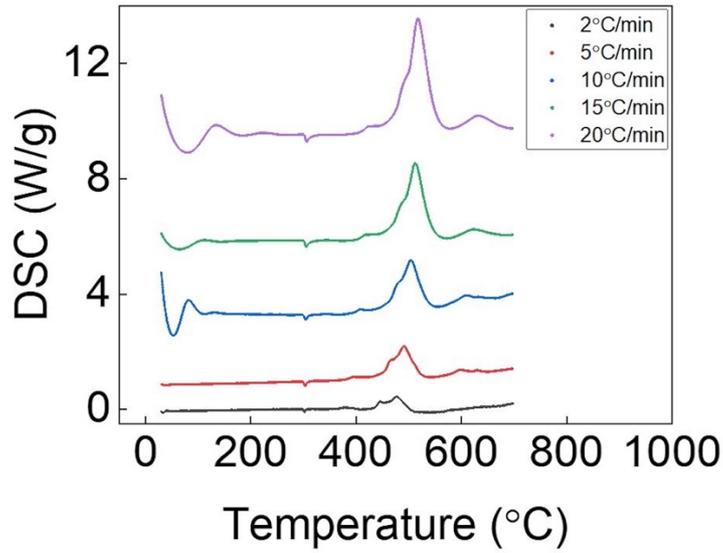

**Fig. 2.** Multiple heating rates DSC curves of WKB mixture (2, 5, 10, 15, 20°C /min).

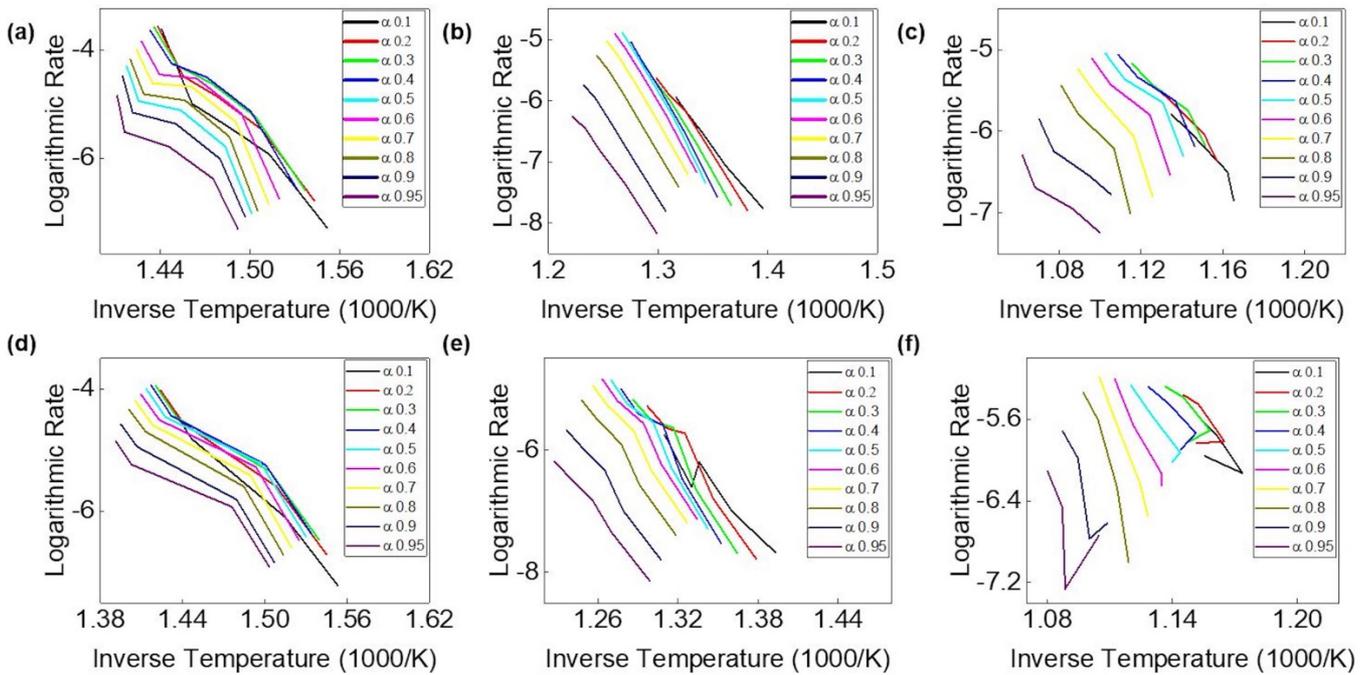

**Fig. 3.** 1/T vs. ln(dα/dt) plots for (a)-(c) combustion peaks of WKB mixture Type A (Peak 1, 2, 3) and (d)-(f) combustion peaks of WKB mixture Type B (Peak 1, 2, 3).

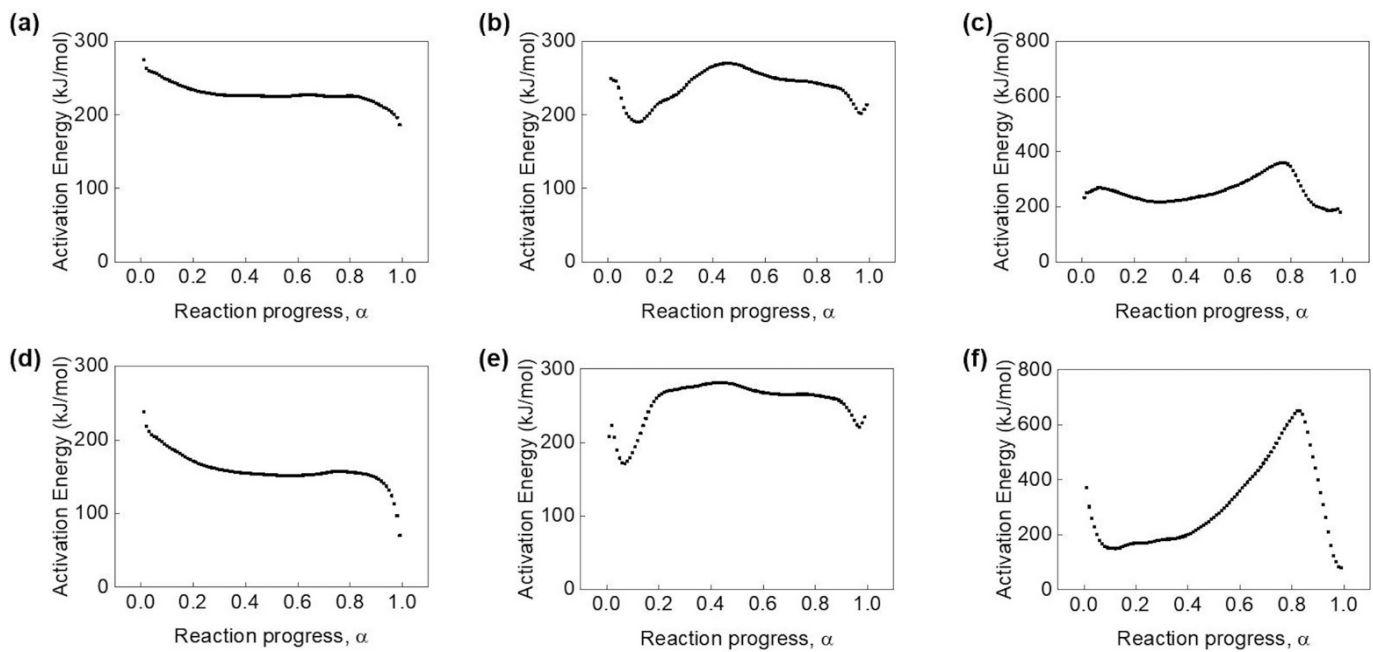

**Fig. 4.** Activation energy vs. reaction progress plots for (a)-(c) combustion peaks of WKB mixture Type A (Peak 1, 2, 3) and (d)-(f) combustion peaks of WKB mixture Type B (Peak 1, 2, 3).

To properly characterize the combustion characteristics of a mixture, it is necessary to interpret each step comprising a reaction by their individual components of time and temperature. Such data may be obtained via a multiple heating rates DSC experiment, where the shifting effects of heating-rate on individual peaks are easy to recognize. The results are presented in Fig. 2, where we observed that some of the combustion peaks that are detectable in most heating rates were not present in some heating rates. Specifically, Peak 3 disappeared in the DSC data measured at the heating rate of 2°C/min. For such reasons, only four points were used to track the variation in α for peak 3, as opposed to five, which was the case for peaks 1 and 2. Despite the partial void in the dataset, however, we were able to extract the necessary kinetic parameters from the existing data.

Aside from the leftmost endothermic peak which corresponds to the solidus-to-solidus phase transition of $KClO_4$, we were able to associate the three remaining peaks of Fig. 2 with the exothermic peaks identified in Fig. 1(d). With this in mind, we investigated the reaction kinetics of the three peaks by following the steps highlighted in part II and III of the Experimental Section. We computed the reaction rate and the temperature for fixed values of α from 0.01 to 0.99 with 0.01 increments. This was repeated for the DSC data measured at all heating rates to amass a collection of 1/T vs. ln (dα/dt) plots for each combustion peak of both mixture types, which gives the schematics of Fig. 3. For simplistic representation, data only corresponding to selected values of α, which are 0.1 to 0.9 with 0.1 increments and a final value of 0.95, is shown.

The overall tendency of the logarithmic reaction rate seems to correspond well across the disparate mixture types, showing that the pyrotechnic behavior of DSC peaks identified in the initial TG-DSC analysis (Fig. 1) presents a coherent trend in the resulting data of the respective peaks. To obtain a greater insight into the reaction kinetics, this was followed up by individually interpolating the plots for the different values of α and extending them to obtain the slope and y-intercept. When multiplied by R, the slope gives the activation energy as a function of reaction progress α as given in Fig. 4, which in itself implicates the change that takes place in the kinetic landscape as the combustion reaction proceeds.

### 3.3 Visualization of the Reaction Dynamics

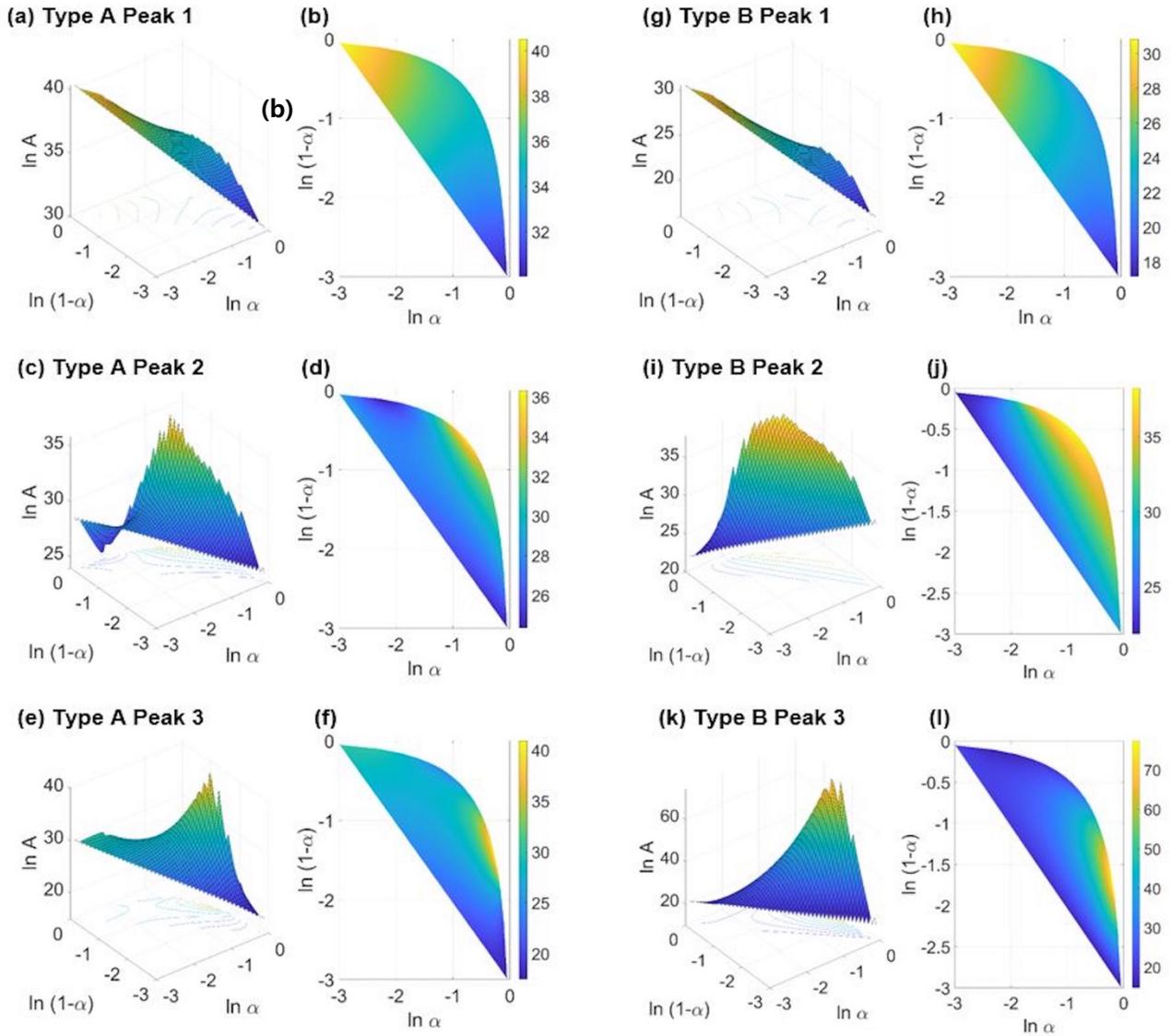

**Fig. 5.** Color gradient plot of the collision factor along the curved line in the ln(1- α) vs. ln α axes plane. (a)-(f), WKB Mixture Type A, (g)-(l), WKB Mixture Type B.

**Table 1.** Kinetic variables and error: m,n, ln(A) and R-squared.

|  | m | n | ln(A) | R-squared |
| --- | --- | --- | --- | --- |
| Type A Peak 1 | -1.4379 | 1.4124 | 35.3303 | 0.91 |
| Type A Peak 2 | 0.6281 | 0.1836 | 31.0995 | 0.254 |
| Type A Peak 3 | 0.2104 | 0.3196 | 28.7858 | 0.0759 |
| Type B Peak 1 | -1.9581 | 1.7383 | 23.0726 | 0.841 |
| Type B Peak 2 | 1.3688 | -0.2015 | 33.9845 | 0.217 |
| Type B Peak 3 | 8.8131 | -7.7540 | 35.4940 | 0.361 |

The y-intercept of the linear interpolation was used to design a color gradient plot of the logarithmic pre-exponential factor along a curved line in the ln(1–α) vs. ln α axes plane. The computed values of reaction parameters m and n, the logarithmic pre-exponential factor, and the least squares error corresponding to each peak of the mixtures are given in Fig. 5. Notice how the x and y-axis values are correlated as implicit functions, implicating that the reaction proceeds along the curved axis of the colored region [Fig. 5(b), (d), (f), (h), (j), (l)] in the clockwise direction.

The degree of reaction-favorable interaction that occur in each point of the individual reaction is represented by the logarithmic pre-exponential factor, whose increase is represented by the color variation of the schematics from blue to yellow. While the reaction parameters m and n do not provide a quantitative measure of the reaction in question, it provides an insight into the nature of the reaction taking place, when substantiated by our evaluation of the activation energy and the intuitive visualization as presented by the color gradient plot. By paying attention to the polarity of the reaction parameters, we can apply that if m is positive, the reaction is nucleation-dominant. If n is positive, the reaction is collision-dominant. The combustion peaks demonstrate that several different combinations of m/n polarities are possible, which, when taken together with the actual kinetic quantities such as the activation energy, suggest a new pathway in administering the reaction rate crucial in controlling the delay time of combustion WKB mixtures.

## 4. Conclusions

WKB mixture is a commonly used compound for manipulating the ignition time, frequented with combustion delay mixtures in military operations. Despite the long history of scrutiny on its chemical properties and various experimental conditions that affect its combustion, work remained unsolved in determining the kinetic parameters involved in the reaction, due to the complexity of the system. To this end, we devised an isoconversional method of measuring the activation energy as a function of the reaction progress for reactions proceeding at disparate temperature ranges, based on data extracted from a conventional multi-heating rates TG-DSC measurement. The same data was used to create a color gradient plot of the logarithmic pre-exponential factor along a curved reaction axis, to fit the Sestak-Berggren solid-state reaction model. Our results present a new, inventive strategy in analyzing the combustion dynamics of the WKB delay mixture, and by extension, any type of combustion delay mixtures.

## 5. Declaration of interests

The authors declare that they have no known competing financial interests or personal relationships that could have appeared to influence the work reported in this paper.

## 6. Acknowledgements

This research was supported by "Poongsan-KAIST Future Research Center Project" and the authors gratefully acknowledge the financial support of Poongsan Corporation.